\documentclass[pdflatex,sn-basic]{sn-jnl}

\usepackage{graphicx}%
\usepackage{multirow}%
\usepackage{amsmath,amssymb,amsfonts}%
\usepackage{amsthm}%
\usepackage{mathrsfs}%
\usepackage[title]{appendix}%
\usepackage{xcolor}%
\usepackage{textcomp}%
\usepackage{manyfoot}%
\usepackage{booktabs}%
\usepackage{algorithm}%
\usepackage{algorithmicx}%
\usepackage{algpseudocode}%
\usepackage{listings}%

\usepackage{natbib, color, setspace, amstext, booktabs, fancyhdr, multirow, float, bm, lineno, xr, subfigure, enumerate, comment} 

\allowdisplaybreaks 
\usepackage{pdflscape}
\usepackage{placeins} 
\usepackage{longtable}
\usepackage[parfill]{parskip}

\newcommand*\patchAmsMathEnvironmentForLineno[1]{%
  \expandafter\let\csname old#1\expandafter\endcsname\csname #1\endcsname
  \expandafter\let\csname oldend#1\expandafter\endcsname\csname end#1\endcsname
  \renewenvironment{#1}%
     {\linenomath\csname old#1\endcsname}%
     {\csname oldend#1\endcsname\endlinenomath}}%
\newcommand*\patchBothAmsMathEnvironmentsForLineno[1]{%
  \patchAmsMathEnvironmentForLineno{#1}%
  \patchAmsMathEnvironmentForLineno{#1*}}%
\AtBeginDocument{%
\patchBothAmsMathEnvironmentsForLineno{equation}%
\patchBothAmsMathEnvironmentsForLineno{align}%
\patchBothAmsMathEnvironmentsForLineno{flalign}%
\patchBothAmsMathEnvironmentsForLineno{alignat}%
\patchBothAmsMathEnvironmentsForLineno{gather}%
\patchBothAmsMathEnvironmentsForLineno{multline}%
}

\usepackage{etoolbox}
\usepackage{tikz}
\usepackage{array}

\theoremstyle{thmstyleone}%
\theoremstyle{thmstyletwo}%

\theoremstyle{thmstylethree}%

\def\bmbeta{\bm{\beta}}

\def\bmSigma{\bm{\Sigma}}
\def\bmalpha{\bm{\alpha}}

\usepackage{hyperref}
\hypersetup{colorlinks = true, linkcolor = blue, urlcolor = blue, citecolor = blue}


\newcounter{parentnumber}

\begin{document}

\title[Bias-Adjusted Attribution Estimation for Rainfall Enhancement Trials]{Bias-Adjusted Attribution Estimation for Rainfall Enhancement Trials}


\author*[1]{\fnm{Zhi Yang} \sur{Tho}}\email{ZhiYang.Tho@anu.edu.au}

\author[1]{\fnm{Raymond} \sur{Chambers}}\email{Raymond.Chambers@anu.edu.au}

\author[1]{\fnm{A.H.} \sur{Welsh}}\email{Alan.Welsh@anu.edu.au}

\affil[1]{\orgdiv{Research School of Finance, Actuarial Studies and Statistics}, \orgname{The Australian National University}, \orgaddress{\street{26C Kingsley Street}, \city{Acton}, \postcode{2601}, \state{ACT}, \country{Australia}}}




\abstract{Model-based analyses of rainfall enhancement trial data typically involve modelling log-transformed rainfall using linear mixed models to assess the effectiveness of enhancement methods under real-world conditions. This approach improves on traditional average-based analyses by allowing explicit control for the effects of meteorological and topographical covariates that may affect precipitation amounts. However, a key issue with such analyses is the bias that arises when back-transforming the log-rainfall to the original scale for estimating attribution, defined as the additional raw-scale rainfall attributable to the enhancement method. To address this issue, we propose a new attribution estimator that incorporates theoretically justified, observation-specific bias-adjustment terms. The proposed estimator improves upon existing estimators that rely on arbitrary adjustments, and satisfies a coherence property that ensures zero estimated attribution for observations without enhancement intervention. A proportional random effect block bootstrap is further used to conduct inference on the attribution quantities. Applying both the proposed estimator and an existing estimator to the Oman rainfall enhancement trial from 2013 to 2018, we find statistically significant positive effect of the ground-based ionization technology on downwind rainfall at the 5\% significance level, with our proposed estimator indicating a smaller effect than the existing method. A simulation study further support the findings based on the proposed estimator, demonstrating its superior estimation accuracy and improved inferential performance of the associated bootstrap confidence intervals.}




\maketitle

\section{Introduction}\label{sec:intro}
Water is one of our most essential resources, as it underpins everything from energy production and industrial processes to agriculture. A major source of water is rainfall, yet climate change has contributed to a concerning decline in rainfall, creating significant challenges for societies and ecosystems around the world. These growing pressures have led to increasing interest in weather-modification methods as a potential way to address water scarcity. 

Rainfall enhancement is a weather modification strategy that aims to increase the amount of precipitation (rain or snow) that falls from clouds, and is typically used in regions facing water shortages or drought. 
An important note regarding rainfall enhancement methods is that they are used to `enhance' precipitation, but not `create' precipitation, so the clouds must already have enough moisture. Cloud seeding has been employed for over seventy years, with particles such as silver iodide or salt released by aircraft or ground-based generators to try to stimulate the formation of raindrops or ice crystals, and ultimately enhance precipitation \citep[e.g.,][]{mantonANDwarren2011, mantonETAL2011,mantonETAL2017, breedETAL2014, rasmussenETAL2018,frenchETAL2018, tessendorfETAL2018}. More recently, unmanned aerial vehicles (UAVs) have begun to replace aircraft for cloud seeding, because they offer greater precision and flexibility in targeting suitable clouds \citep[e.g.,][]{jungETAL2022,defeliceETAL2023,millerETAL2024}. However, cloud seeding has faced criticism due to its high operational costs and uncertainties about environmental impacts \citep{cooperANDjolly1970,kleinANDmolise1975,fajardoETAL2016}. To address some of these limitations, ground-based or UAV-based ionization technologies have emerged as alternative strategies.  They use charged ions to try to stimulate the coalescence rate of cloud droplets and subsequently increase rainfall \citep{beareETAL2010, beareETAL2011, zhengETAL2020, harrisonETAL2021}.

Despite conclusive evidence from controlled laboratory experiments that releasing silver iodide or charged ions can accelerate water droplet formation \citep[see e.g.,][]{demott1995,hortalETAL2012,maETAL2020}, their effectiveness in enhancing rainfall at a real-world scale remains uncertain. Field trials of cloud seeding and ground-based ionization are conducted in natural environments where full experimental control is infeasible, and are therefore more appropriately viewed as observational studies subject to atmospheric variability and confounding factors. Statistical analyses of rainfall enhancement trials are thus required to assess the effectiveness of different enhancement methods under real-world conditions. A commonly used approach is averaged-based analysis, which compares average rainfall observed in the trial area during the trial period with either average rainfall in surrounding non-trial areas over the same period, or average rainfall in the same trial area outside the trial period \citep[e.g.,][]{morrisonETAL2009, levinETAL2010,zhengETAL2021}. However, this method does not explicitly account for the effects of meteorological and topographical variables, and may yield biased or misleading conclusions when these conditions differ across areas or time periods.

To overcome this limitation, \cite{chambersETAL2012,chambersETAL2016,chambersETAL2022} proposed a model-based analysis that fits a linear mixed model to gauge-day level precipitation data. This framework allows for explicit control of variables that affect precipitation amounts, e.g., meteorological and topographical variables. Rainfall values were log-transformed to better satisfy the normality assumption of the linear mixed model. Leveraging the parametric structure of the model, \cite{beareETAL2010} and \cite{chambersETAL2022} decomposed the observed rainfall values into two components: natural rainfall (i.e., the rainfall that would have occured in the absence of any rainfall enhancement intervention) and the rainfall enhancement effect. Based on this decomposition, they defined attribution quantities that can be interpreted as the amount of rainfall, on the raw-scale, attributable to the operation of the rainfall enhancement technology, where the focus on raw-scale rainfall aligns with the guidelines set out by the World Meteorological Organization \citep{wmo2010}. Estimation of these attribution quantities was further studied by \cite{chambersETAL2022}, who employed a smearing-type adjustment \citep{duan1983} to account for transformation bias arising from back-transforming log-rainfall to the original scale. However, this adjustment relied on arbitrary choices regarding how the adjustment was weighted between the natural rainfall and enhancement effect components. 

Building upon the work of \cite{chambersETAL2022}, we propose an improved estimator for the attribution quantities, obtained by appropriately correcting for the aforementioned transformation bias using the estimated variance-covariance matrix of the estimator of the fixed effect coefficients. This proposed estimator offers several advantages over the approach studied by \cite{chambersETAL2022}. First, the adjustment that we propose is theoretically grounded and formally justified. Second, it avoids any arbitrary weighting between the adjustments applied to the natural rainfall and enhancement effect components. Third, our estimated attribution quantities are always zero for observations that did not receive any rainfall enhancement intervention, a property that is not guaranteed by the estimator of \cite{chambersETAL2022}.

We demonstrate theoretically the transformation bias that arises from using unadjusted estimators of the attribution quantities and show that the proposed attribution estimators effectively eliminate this bias. In addition, we adopt a semiparametric bootstrap method recently proposed by \cite{thoETAL2025} to quantify uncertainty in the proposed attribution estimates, thereby satisfying the \cite{wmo2010} requirement for proper uncertainty quantification of estimated rainfall increases attributable to rainfall enhancement technology. Both the proposed attribution estimator and the estimator of \cite{chambersETAL2022} are applied to rainfall enhancement trial data from Oman spanning 2013-2018 to assess the effectiveness of the ground-based ionization technology. This five-year long trial involved up to ten ionizers and 201 rain gauges under a random cross-over design, in which half of the ionizers were randomly activated and the remainder deactivated on each trial day. The results indicate that the proposed attribution estimates yield a smaller magnitude of enhancement effect than those of \cite{chambersETAL2022}, although both approaches provide evidence of a statistically significant positive enhancement effect at the 5\% significance level. To further investigate the discrepancy in magnitude between the two estimators, we conduct a simulation study designed to mimic the Oman trial setting. The simulation results show that the proposed attribution estimator outperforms \cite{chambersETAL2022}'s estimator in recovering the true attribution quantities, and that the bootstrap method achieves satisfactory coverage when paired with the proposed attribution estimator. 

The remainder of this article is organized as follows. Section \ref{sec:trial} describes the motivating dataset, the Oman rainfall enhancement trial.  Section \ref{sec:model} introduces the linear mixed model and defines the attribution quantities of interest. Section \ref{sec:bias_adjustment_attribution} presents the proposed attribution estimator, establishes its unbiasedness property, compares it with the existing estimator, and outlines a bootstrap procedure for inference. Section \ref{sec:realdata} applies both the proposed and existing attribution estimators to the Oman trial data. Section \ref{sec:simulation} reports the results of simulation studies, and Section \ref{sec:conclusion} concludes with some final remarks. 

\section{The Oman 2013-2018 Trial}\label{sec:trial}

This trial was conducted in the Hajar Mountains region of Oman and investigated a ground-based ionization technology known as Atlant. The Atlant system uses a high-voltage power supply to generate corona discharges from a wire array, referred to as an ionizer, with the aim of negatively charging the surrounding aerosols. The hope is that these ionized aerosols are subsequently transported by atmospheric uplift to cloud layers downwind of the ionizer, where they act as cloud condensation nuclei (CCN). This process, known as cloud ionization, is hypothesized to enhance the collision-coaslescence rate of cloud droplets, thereby promoting raindrop formation and potentially increasing the amount of rainfall observed downwind of the ionizer. 

\begin{table}[!tb]
\caption{Number of ionizers and rain gauges in each year of the Oman trial.}
\label{table:no_ionizers_gauges}
\centering
\begin{tabular}{l rrrrrr}
\toprule
& \multicolumn{6}{c}{Year}  \\
 \cmidrule(l){2-7}
& 2013 & 2014 & 2015 & 2016 & 2017 & 2018 \\
\midrule
Ionizer & 2 & 4 & 6 & 8 & 10 & 10\\
Rain Gauge & 120 & 149 & 179 & 191 & 201 & 201 \\ 
\bottomrule
\end{tabular}
\end{table}

\begin{figure}[!tb]
    \centering
    \includegraphics[width=1\linewidth]{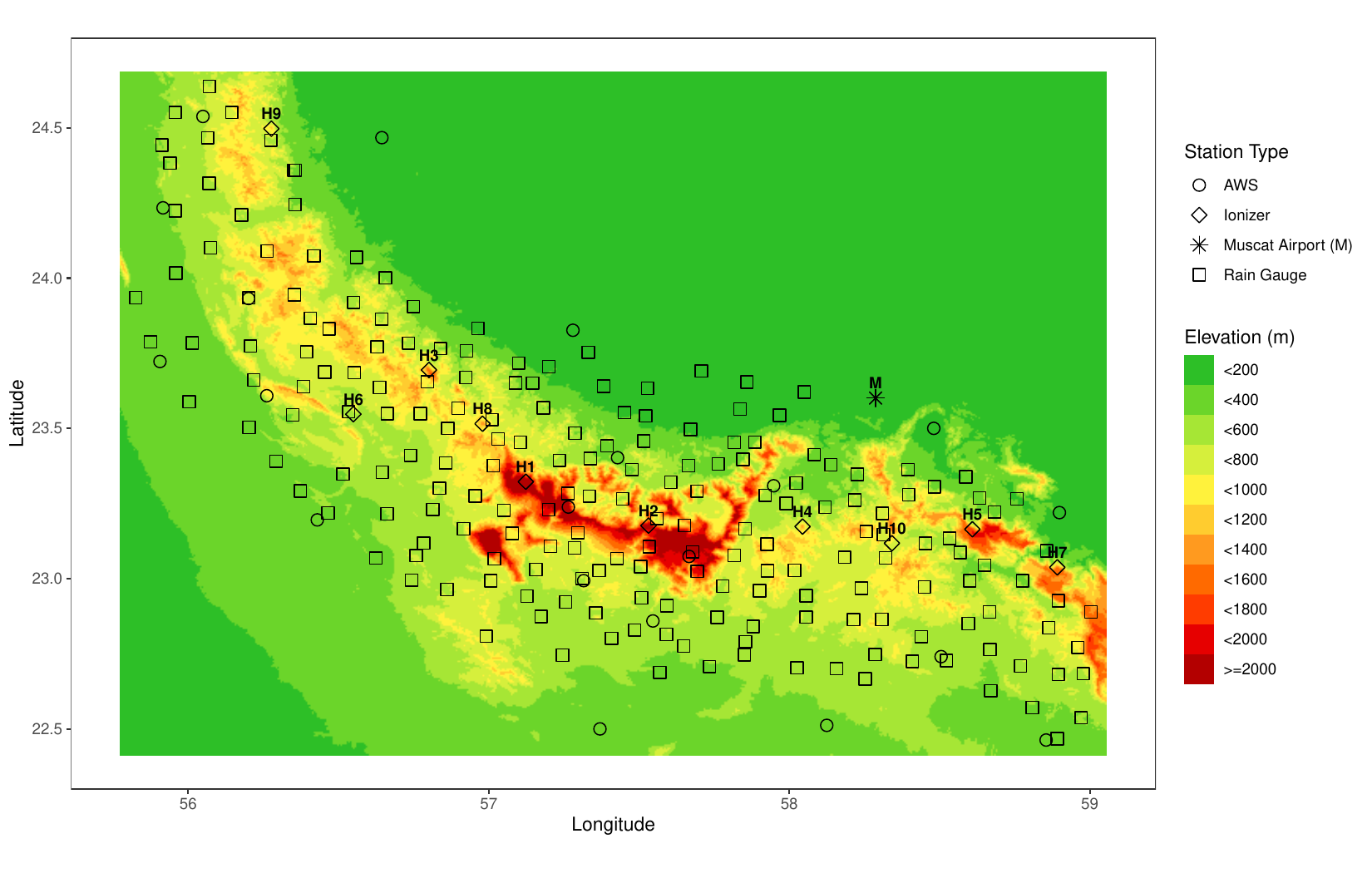}
    \caption{Location of ionizers (diamonds), rain gauges (squares), AWS (circles), and Muscat International Airport (M) across Hajar Mountains region of Oman, shown on a contour map of elevation.}
    \label{fig:map_stations}
\end{figure}

The trial began in 2013 with two ionizers and expanded to a total of ten ionizers (denoted H1 - H10) by 2018. In addition to the ionizers, a comprehensive network of rain gauges was installed throughout the trial area, providing daily rainfall measurements for all $488$ trial days. The network initially consisted of 120 rain gauges in 2013, arranged on an approximately 10km regular grid across the trial area, and expanded to 201 rain gauges in 2018 to accommodate the enlargement of the trial area associated with the deployment of additional ionizers. Table \ref{table:no_ionizers_gauges} summarizes the number of ionizers and rain gauges in each year. The trial also incorporated meteorological data from automatic weather stations (AWS) operated by Oman Directorate-General of Meteorology and Aerial Navigation (DGMAN) in the Hajar Mountains region. These data include daily averages of wind speed, dry air temperature, relative humidity and air pressure at AWS ground level. In addition, daily DGMAN radiosonde observations at Muscat International Airport were used to obtain vertical wind profiles along with daily atmospheric indicators related to storm development. 
Figure \ref{fig:map_stations} shows the locations of the ionizers, rain gauges, AWS, and Muscat International Airport within the Hajar Mountains region.

A randomized cross-over design was adopted for the trial, whereby, on each trial day, half of the deployed ionizers were switched on while the remaining half were switched off according to a randomized operating schedule specified prior to the commencement of the trial each year. As the working hypothesis underyling the ionization technology relies on the downwind transport of ionized aerosols, \cite{chambersETAL2016} defined the downwind region of an ionizer as a rectangular corridor extending 75km in length and 30km in width in the daily steering wind direction from the ionizer. The daily steering wind direction was computed using radiosonde data from Muscat International Airport. This definition results in dynamic downwind corridors, since the steering wind direction varies from day to day. Consequently, downwind gauge-day observations (which constitute the focus of the trial) are defined as those corresponding to rain gauges located within at least one such downwind corridor of all deployed ionizers on a given trial day. Among these observations, gauges that are downwind of at least one active ionizer are classified as `target' observations that receive the enhancement intervention, whereas those that are downwind only of inactive ionizers are classified as `control' observations that are not affected by ionization.

\begin{figure}[!tb]
    \centering
    \includegraphics[width=1\linewidth]{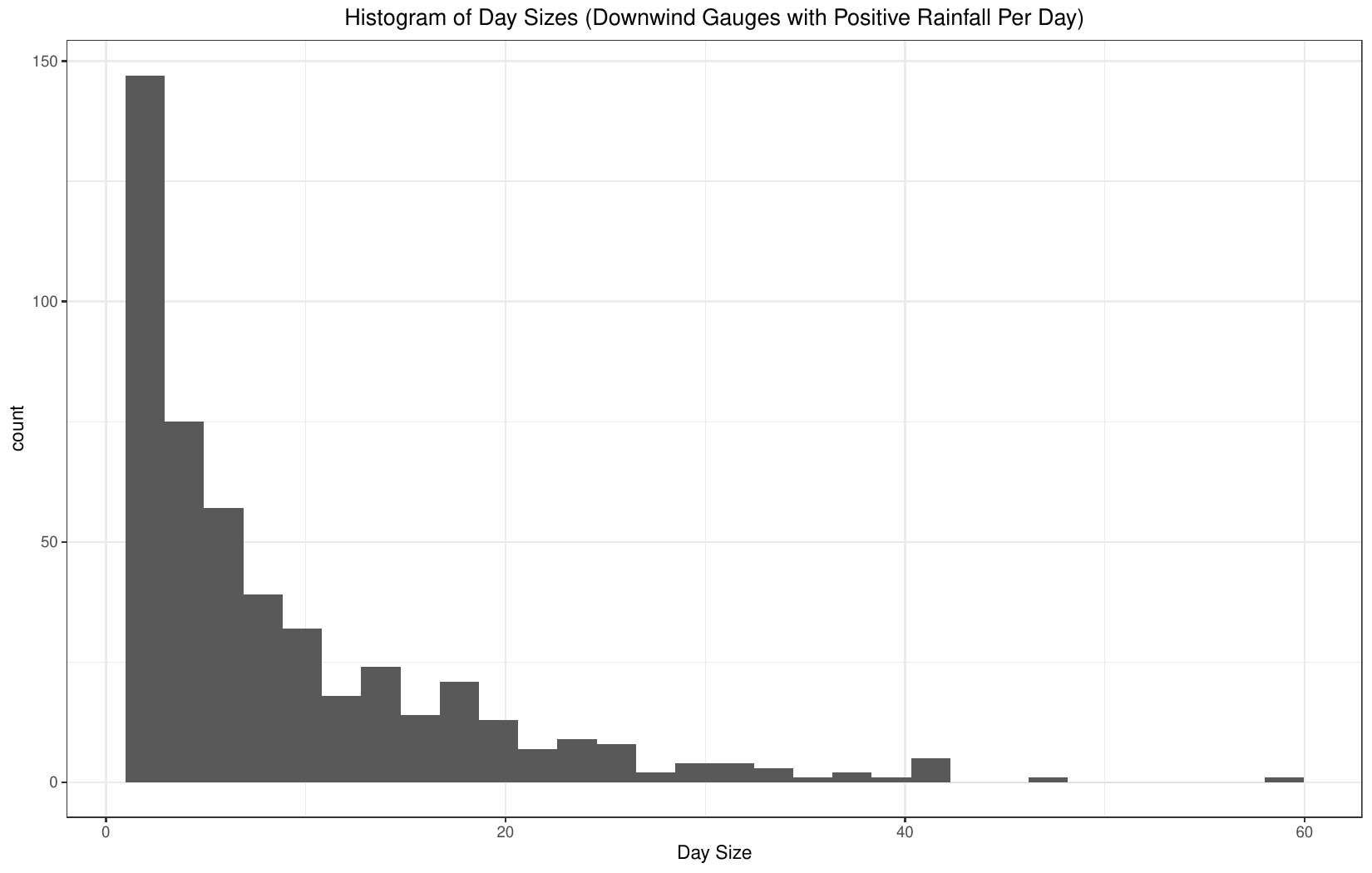}
    \caption{Histogram of day sizes (number of downwind rain gauges with positive rainfall) in the Oman trial.}
    \label{fig:hist_nt}
\end{figure}

\begin{figure}[!tb]
    \centering
    \includegraphics[width=1\linewidth]{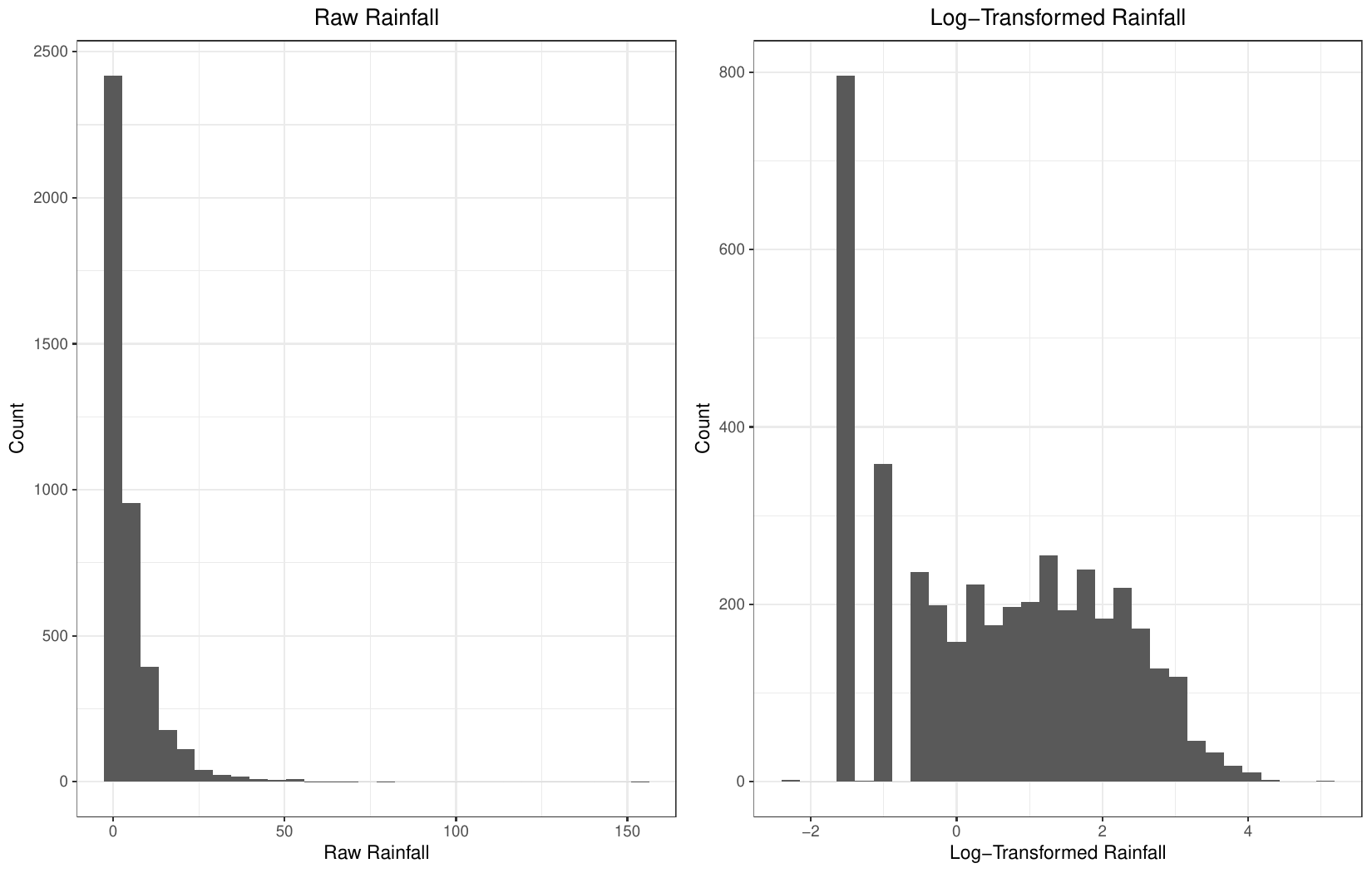}
    \caption{Histograms of raw rainfall (left) and log-transformed rainfall (right) in the Oman trial.}
    \label{fig:hist_raw_log_rainfall}
\end{figure}

As discussed in Section \ref{sec:intro}, rainfall enhancement methods, including the Atlant ionization technology, are designed to enhance rainfall rather than create rainfall. Consequently, the analysis focuses on downwind gauge-day observations with positive rainfall amounts. This results in a total of $4168$ gauge-day observations, comprising 2176 target observations and 1992 control observations. Figure \ref{fig:hist_nt} presents the distribution of day sizes, defined as the number of downwind observations with positive rainfall on each trial day. It shows that the day sizes are highly unbalanced, with most of the days having fewer than 20 downwind gauges with positive rainfall, while a small number of days have substantially larger number of such gauges, reflecting the inherent variability of rainfall. 
Figure \ref{fig:hist_raw_log_rainfall} further shows the distributions of rainfall on both the raw and logarithmic scales, indicating that the raw rainfall is heavily right-skewed, whereas log-rainfall is considerably more symmetric and closer to a normal distribution. 

\section{Rainfall Model and Attribution} \label{sec:model}
We focus on the study of gauge-day rainfall enhancement data as measured in the Oman trial. Let $y_{it} > 0$ denote the observed positive rainfall at rain gauge $i$ on trial day $t$ for $i= 1,\cdots,n_t$ and $t = 1,\cdots,T$ with $T = 488$ days, where $n_t$ denotes the number of rain gauges that record positive rainfall on trial day $t$ and $N = \sum_{t=1}^{T} n_t = 4168$ observations. 
We follow \cite{chambersETAL2012,chambersETAL2016,chambersETAL2022} to model the log-transformed rainfall using a linear mixed model:
\begin{equation}
    \log(y_{it}) = \bm{x}_{it}^\top \bmalpha + \bm{z}_{it}^\top \bmbeta + u_t + e_{it},
    \label{eq:lograin_lmm}
\end{equation}
where $\bm{x}_{it}$ is a $p$-dimensional vector containing covariates that can affect the amount of observed rainfall and are unrelated to rainfall enhancement intervention, e.g., meteorological and topographical covariates, $\bm{z}_{it}$ is a $q$-dimensional vector containing covariates that quantify the exposure to rainfall enhancement intervention, $\bmalpha$ and $\bmbeta$ are fixed effect coefficients that capture the effect of their corresponding covariates on rainfall, $u_t$ are independent and identically distributed (i.i.d.) day random effects with mean $0$ and variance $\sigma^2_u$, and $e_{it}$ are i.i.d. error terms with mean $0$ and homogeneous variance $\sigma^2_e$. Importantly, the random effects $u_t$ and error terms $e_{it}$ are assumed to follow Gaussian distributions \citep{pinheiroANDbates2000,batesETAL2015}. This motivates modelling log-transformed rainfall $\log(y_{it})$ rather than the raw scale $y_{it}$, as the latter is often right-skewed and exhibits heteroskedasticity, with variability increasing with the mean. 
In contrast, modelling rainfall on the logarithmic scale tends to stabilize the variance, thereby aligning better with the homoscedastic error assumption, and log-rainfall is often much closer to being symmetric and bell-shaped; see Figure \ref{fig:hist_raw_log_rainfall}. 

The inclusion of covariates $\bm{x}_{it}$ in model \eqref{eq:lograin_lmm} allows potential confounding effects on rainfall to be controlled for, which is an advantage over traditional averaged-based analyses. For example, incorporating topographical variables into $\bm{x}_{it}$ enables the model to account for systematic differences in rainfall across locations with varying topographical characteristics. 
Meanwhile, the day random effects $u_t$ are introduced to capture residual temporal heterogeneity in rainfall that is not explained by the covariates. The parameters $(\bmalpha^\top, \bmbeta^\top, \sigma^2_u, \sigma^2_e)^\top$ can be estimated using the maximum likelihood (ML) or restricted maximum likelihood (REML) approaches \citep{patternsonANDthompson1971,harville1977}, which we denote as $(\hat{\bmalpha}^\top, \hat{\bmbeta}^\top, \hat{\sigma}^2_u, \hat{\sigma}^2_e)^\top$. 

The key research question in a rainfall enhancement trial is to quantify how much of the observed positive rainfall can be attributed to the intervention of the rainfall enhancement method. As noted in Section \ref{sec:intro}, this change in rainfall should be expressed on the raw scale for interpretability \citep{wmo2010}, and is commonly referred to as attributed rainfall, or simply attribution. We formally define the attribution at gauge $i$ on day $t$ as
\begin{equation}
    A_{it} = y_{it} - R_{it},
    \label{eq:Ait}
\end{equation}
where $R_{it}$ is the latent natural rainfall that would have been observed in the absence of any rainfall enhancement intervention.  These attributions can be summed over all gauge-day observations and expressed relative to total natural rainfall, i.e., $$\theta = \frac{\sum_{t=1}^{T} \sum_{i=1}^{n_t} A_{it}}{\sum_{t=1}^{T} \sum_{i=1}^{n_t} R_{it}}\times 100\%,$$ giving the overall attribution 
as a percentage of the total natural rainfall. Here, $\theta$ is the parameter of interest in this article. 

To define the unobserved $R_{it}$, \cite{beareETAL2010} and \cite{chambersETAL2022} decomposed the observed rainfall into
\begin{equation*}
    y_{it} = R_{it} E_{it},
\end{equation*}
where $R_{it}$ is the latent natural rainfall and $E_{it} > 0$ represents the multiplicative enhancement effect. Values of $E_{it} > 1$ ($E_{it}< 1$) correspond to an increase (decrease) in rainfall due to the intervention. 
Since model \eqref{eq:lograin_lmm} can be written as $y_{it}= \exp(\bm{x}_{it}^\top \bmalpha +\bm{z}_{it}^\top \bmbeta + u_t + e_{it})$, and $\bm{z}_{it}$ measures exposure to enhancement while $\bm{x}_{it}$ is unrelated to the intervention, we set $E_{it} = \exp(\bm{z}_{it}^\top \bmbeta)$ and 
\begin{equation}
    R_{it} = \exp(\bm{x}_{it}^\top \bmalpha + u_t + e_{it}) = y_{it} \exp(-\bm{z}_{it}^\top \bmbeta).
    \label{eq:Rit}
\end{equation}
As the attribution $A_{it}$ (and hence $\theta$) depends on the unobserved $R_{it}$, we propose corresponding estimates in the next section, which properly account for the transformation bias arising from the log-transformation.

\section{Bias-Adjusted Attribution Estimates} \label{sec:bias_adjustment_attribution}
We can estimate the latent natural rainfall $R_{it}$ by replacing $\exp(-\bm{z}_{it}^\top \bmbeta)$  with $\exp(-\bm{z}_{it}^\top \hat{\bmbeta})$ in equation \eqref{eq:Rit}. Under mild conditions, the fixed effect estimators are asymptotically normal, i.e., $\hat{\bmbeta} \stackrel{a}{\sim} N(\bmbeta, \bm{\Sigma}_{\bmbeta})$ with $\bm{\Sigma}_{\bmbeta} = \{\sum_{t=1}^{T} \bm{Z}_t^\top \bmSigma_t^{-1}(\sigma^2_u, \sigma^2_e) \bm{Z}_t\}^{-1} $ \citep{lairdANDware1982}, where $\bm{Z}_t = (\bm{z}_{1t},\cdots,\bm{z}_{n_t t})^\top$,  $\bmSigma_t(\sigma^2_u, \sigma^2_e) = \sigma^2_u \bm{1}_{n_t}\bm{1}_{n_t}^\top + \sigma^2_e \bm{I}_{n_t}$ and $\bm{1}_{n_t}$ is the $n_t$-dimensional vector of ones. It then follows that $\exp(-\bm{z}_{it}^\top \hat{\bmbeta})$ is a biased estimator  of $\exp(-\bm{z}_{it}^\top \bmbeta)$, as
\begin{align}
    \mathrm{E}\{\exp(- \bm{z}_{it}^\top \hat{\bm{\beta}}) \}
    &= \mathrm{E}[\exp(- \bm{z}_{it}^\top \bm{\beta}) \exp\{- \bm{z}_{it}^\top (\hat{\bm{\beta}} - \bm{\beta})\} ] \notag \\
&= \exp(- \bm{z}_{it}^\top \bm{\beta}) \mathrm{E}[\exp\{- \bm{z}_{it}^\top (\hat{\bm{\beta}} - \bm{\beta})\} ]  \notag \\
&= \exp(- \bm{z}_{it}^\top \bm{\beta}) \exp \left\{ \frac{1}{2} \bm{z}_{it}^\top \bm{\Sigma}_{\bmbeta} \bm{z}_{it}\right\}   \neq \exp(-\bm{z}_{it}^\top \bmbeta).
\label{eq:bias_derivation}
\end{align}
The above derivation suggests that an appropriate bias-adjusted estimate of $R_{it}$ could take the form $y_{it} \lambda_{it} \exp(-\bm{z}_{it}^\top \hat{\bmbeta})$ with the adjustment term $\lambda_{it} = \exp \{-(1/2) \bm{z}_{it}^\top \bmSigma_{\bmbeta} \bm{z}_{it} \}$. However, $\bmSigma_{\bmbeta}$ involves the unknown parameters $(\sigma^2_u,\sigma^2_e)$, motivating our proposal of the bias-adjusted estimator 
\begin{equation*}
    \hat{R}_{it} = y_{it} \hat{\lambda}_{it} \exp(-\bm{z}_{it}^\top \hat{\bmbeta}), \textrm{ with } \hat{\lambda}_{it} = \exp\left\{ -\frac{1}{2} \bm{z}_{it}^\top \hat{\bmSigma}_{\bmbeta} \bm{z}_{it} \right\},
\end{equation*}
where $\hat{\bmSigma}_{\bmbeta} = \{\sum_{t=1}^{T} \bm{Z}_t^\top \bmSigma_t^{-1}(\hat{\sigma}^2_u,\hat{\sigma}^2_e) \bm{Z}_t\}^{-1} $ is an estimator of $\bmSigma_{\bmbeta}$.
Subsequently, the attribution can be estimated as
\begin{equation*}
    \hat{A}_{it} = y_{it} - \hat{R}_{it} = y_{it} \left\{ 1 - \hat{\lambda}_{it}\exp(-\bm{z}_{it}^\top \hat{\bmbeta}) \right\},
\end{equation*}
and
\begin{equation*}
    \hat{\theta} = \frac{ \sum_{t=1}^{T} \sum_{i=1}^{n_t} \hat{A}_{it} }{ \sum_{t=1}^{T} \sum_{i=1}^{n_t} \hat{R}_{it}  } \times 100\% = \frac{ \sum_{t=1}^{T} \sum_{i=1}^{n_t} y_{it} \left\{ 1 - \hat{\lambda}_{it}\exp(-\bm{z}_{it}^\top \hat{\bmbeta}) \right\} }{ \sum_{t=1}^{T} \sum_{i=1}^{n_t}  y_{it} \hat{\lambda}_{it} \exp(-\bm{z}_{it}^\top \hat{\bmbeta})  } \times 100\%.
\end{equation*}

It is worth noting that the proposed bias-adjustment term $\hat{\lambda}_{it}$ is gauge-day specific, effectively removes the transformation bias arising from the exponentiation in \eqref{eq:bias_derivation}, and ultimately leads to accurate estimation of the attribution for each gauge-day observation. 
Beyond its theoretical justification, the proposed estimator is also coherent with the attribution framework and the rainfall enhancement trial design, ensuring that observations with no intervention are estimated to have zero attributed rainfall and their latent natural rainfall equal to the observed rainfall. More specifically, gauge-day observations that do not receive any enhancement intervention have $\bm{z}_{it} = \bm{0}_q$, where $\bm{0}_{q}$ is the $q$-dimensional zero vector, and hence their corresponding bias-adjustment terms are $\hat{\lambda}_{it} = \exp\{ - (1/2) \bm{0}_q^\top \hat{\bmSigma}_{\bmbeta} \bm{0}_q \} = 1$ with the corresponding estimates
\begin{equation*}
    \hat{R}_{it} = y_{it} \hat{\lambda}_{it} \exp(- \bm{0}_q^\top \hat{\bmbeta}) = y_{it}, \; \hat{A}_{it} = y_{it} - \hat{R}_{it} = y_{it} - y_{it} = 0,
\end{equation*}
which is a desired property. 

\cite{chambersETAL2022} estimated the latent natural rainfall using a similar form $\tilde{R}_{it} = y_{it} \tilde{\lambda} \exp(-\bm{z}_{it}^\top \hat{\bmbeta}) $, where $\tilde{\lambda}$ is a bias-adjustment term that is common across all gauge-day observations. In particular, they proposed to compute
\begin{equation}
    \tilde{\lambda} = \frac{2m}{\sqrt{(1 + m)^2 + 4(\mu - 1)m} + (m - 1)},
    \label{eq:tilde_lambda}
\end{equation}
where $\mu = N^{-1} \sum_{t=1}^{T} \sum_{i=1}^{n_t} \{ y_{it} / \exp(\bm{x}_{it}^\top \hat{\bmalpha} + \bm{z}_{it}^\top\hat{\bmbeta} + \hat{u}_t ) \} $ is the smearing adjustment factor \citep{duan1983} used to predict the raw scale rainfall via $\tilde{y}_{it} = \mu \exp(\bm{x}_{it}^\top \hat{\bmalpha} + \hat{u}_t) \exp(\bm{z}_{it}^\top\hat{\bmbeta} )$, and $\hat{u}_t$ is the empirical best linear unbiased prediction (EBLUP) of the day random effect. One limitation of equation \eqref{eq:tilde_lambda} is that the tuning parameter $m > 0$ must be specified. This choice implicitly determines how the smearing adjustment $\mu$ is decomposed between the natural rainfall component $\exp(\bm{x}_{it}^\top \hat{\bmalpha} + \hat{u}_t)$ and the enhancement component $\exp(\bm{z}_{it}^\top\hat{\bmbeta} )$; see \cite{chambersETAL2022} for details. Larger values of $m$ allocate relatively more adjustment to the former component, and vice versa. For example, setting $m = 0$ yields $\tilde{\lambda} = \mu^{-1}$, while $m = \infty$ gives $\tilde{\lambda} = 1$. \cite{chambersETAL2022} recommended choosing $m = \hat{\mathrm{V}}( \bm{x}_{it}^\top \hat{\bmalpha} + \hat{u}_t  ) / \hat{\mathrm{V}}( \bm{z}_{it}^\top \hat{\bmbeta})$, where $\hat{\mathrm{V}}$ denotes the empirical variance over all gauge-day observations. Individual and overall attribution are then estimated via $\tilde{A}_{it} = y_{it} - \tilde{R}_{it}$ and $\tilde{\theta} = (\sum_{t=1}^{T} \sum_{i=1}^{n_t} \tilde{A}_{it}) /  (\sum_{t=1}^{T} \sum_{i=1}^{n_t} \tilde{R}_{it}) \times 100\%$. In constrast to this approach, our proposed adjustment $\hat{\lambda}_{it}$ is derived directly from the asymptotic distribution of $\hat{\bmbeta}$ and therefore has a clear theoretical justification, avoiding any arbitrary decomposition of the smearing factor. Moreover, under the recommended choice $m = \hat{\mathrm{V}}( \bm{x}_{it}^\top \hat{\bmalpha} + \hat{u}_t  ) / \hat{\mathrm{V}}( \bm{z}_{it}^\top \hat{\bmbeta})$, it is straightforward to verify that $\tilde{R}_{it} = y_{it} \tilde{\lambda} \exp(-\bm{0}_q^\top \hat{\bmbeta}) \neq y_{it} $ and $\tilde{A}_{it} \neq 0$ for gauge-day observations with no enhancement exposure. As a result, their estimator does not satisfy the coherence property implied by the attribution framework and trial design. This occurs because $\tilde{\lambda}$ is constant across observations, whereas our adjustment $\hat{\lambda}_{it}$ is gauge-day specific. Empirical comparisons between the proposed bias-adjusted estimator and that of \cite{chambersETAL2022} are presented in Sections \ref{sec:realdata} -- \ref{sec:simulation}, further illustrating the advantages of our approach.

\subsection{Proportional Random Effect Block Bootstrap for Inference}\label{sec:PREB}
In this section, we discuss a bootstrap procedure that can be used to conduct statistical inference on the attribution quantity of interest, namely the proportional random effect block (PREB) bootstrap proposed by \cite{thoETAL2025}. Let $\mathrm{SRSWR}(\bm{a},c)$ represent the outcome of $c$ independent draws using simple random sampling (SRS) with replacement from the vector $\bm{a} = (a_1,\cdots,a_T)$ and $\mathrm{PPSWR}(\bm{a}, \bm{b},c)$ represent the outcome of $c$ independent draws using probability-proportional-to-size (PPS) sampling with replacement from the vector $\bm{a}$ with corresponding sizes given by the vector $\bm{b} = (b_1,\cdots,b_T)$. In PPS sampling, the probability that $a_t$ is selected in a single draw is $b_t / \sum_{t'=1}^{T} b_{t'} $. The PREB bootstrap procedure is described in Algorithm \ref{al:PREB}.

\begin{algorithm}[H]
\caption{\footnotesize Algorithm for the PREB bootstrap.}
\label{al:PREB}
\begin{algorithmic}[1]

\Statex \textbf{Input:} Estimated $(\hat{\bmalpha}^\top, \hat{\bmbeta}^\top, \hat{\sigma}_u^2, \hat{\sigma}_e^2)^\top$
\Statex \textbf{Output:} Bootstrap samples $\{\hat{\theta}^{*(b)} : b=1,\ldots,B\}$ with superscript $(b)$ indicating the $b$-th bootstrapped estimate.

\State Compute marginal residuals $\bm{r}_{t} = \log(\bm{y}_t) - \bm{X}_t \hat{\bmalpha} - \bm{Z}_t \hat{\bmbeta}$ for $t = 1,\cdots,T$, where $\bm{X}_t = (\bm{x}_{1t},\cdots,\bm{x}_{n_t t})^\top$. Then, calculate the day-level random effect predictors $\hat{u}_t = n_t^{-1} \bm{1}_{n_t}^\top \bm{r}_t$ and the residuals  $\hat{\bm{e}}_t = \bm{r}_t - \hat{u}_t \bm{1}_{n_t}$ for $t=1,\cdots,T$.

\State Compute reflated day-level random effect predictors $\hat{u}_{t}^{sc} = \hat{u}_t^c \hat{\sigma}_u \{ T^{-1} \sum_{t' = 1}^{T} (\hat{u}_{t'}^c)^2 \}^{-1/2}$ with $\hat{u}_t^c = \hat{u}_t - T^{-1} \sum_{t'=1}^{T} \hat{u}_{t'} $, and reflated residuals $\hat{\bm{e}}_t^s  = \hat{\bm{e}}_t \hat{\sigma}_e ( N^{-1} \sum_{t'=1}^{T} \sum_{i'=1}^{n_{t'}} \hat{e}_{i't'}^2 )^{-1/2}$ for $t=1,\cdots,T$.

\For{$b=1,\ldots,B$}
    \State Obtain bootstrap samples of the day-level random effects via  $u_t^{*} = \mathrm{SRSWR}( ( \hat{u}_1^{sc},\cdots,\hat{u}_T^{sc} ),1)$ for $t=1,\cdots,T$.

    \State Obtain bootstrap samples of the unit-level residuals by first sampling the donor cluster $d_t^{*} = \mathrm{PPSWR}( (1,\cdots,T), (n_1,\cdots,n_T) ,1)$, and then sampling $\bm{e}_{t}^{*} 
    = \mathrm{SRSWR}( \hat{\bm{e}}_{d_t^{*}}^s , n_t) $ for $t=1,\cdots,T$ .
    
    \State Obtain bootstrap samples of the log-rainfall as $\log(\bm{y}^{*}_t) = \bm{X}_t \hat{\bmalpha} + \bm{Z}_t \hat{\bmbeta} + u_t^{*} \bm{1}_{n_t} + \bm{e}_t^{*}$ for $t=1,\cdots,T$.

    \State Fit the linear mixed model \eqref{eq:lograin_lmm} to the bootstrap sample data $\{ (\bm{y_t}^{*} , \bm{X}_t, \bm{Z}_t): 1=t,\cdots,T\}$ and obtain bootstrap parameter estimates $(\hat{\bmalpha}^{*\top}, \hat{\bmbeta}^{*\top}, \hat{\sigma}_u^{2*}, \hat{\sigma}_e^{2*})^\top$.

    \State Compute the bias-adjustment term $\hat{\lambda}_{it}^{*} = \exp \{ -(1/2) \bm{z}_{it}^\top \hat{\bmSigma}_{\bmbeta}^{*} \bm{z}_{it} \}$, where $\hat{\bmSigma}_{\bmbeta}^{*(b)} = \{ \sum_{t=1}^{T} \bm{Z}_t^\top \bmSigma_t^{-1}( \hat{\sigma}_u^{2*}, \hat{\sigma}_e^{2*}  ) \bm{Z}_t \}^{-1}$. Then, compute the estimated $\hat{R}_{it}^{*} = y_{it}^{*} \hat{\lambda}_{it}^{*} \exp(-\bm{z}_{it}^\top \hat{\bmbeta}^{*}) $ and estimated individual and overall attribution $\hat{A}_{it}^{*} = y_{it}^{*} - \hat{R}_{it}^{*}$ and $\hat{\theta}^{*} = (\sum_{t=1}^{T} \sum_{i=1}^{n_t} \hat{A}_{it}^{*}) / (\sum_{t=1}^{T} \sum_{i=1}^{n_t} \hat{R}_{it}^{*}) \times 100\%$.
\EndFor

\end{algorithmic}
\end{algorithm}

By repeating the bootstrap a sufficiently large number of times, the bootstrap distribution $\{\hat{\theta}^{*(b)}  : b = 1,\cdots, B\}$ allows the construction of bootstrap confidence intervals for the overall attribution $\theta$. For example, a $100(1-\alpha)\%$ confidence interval for $\theta$ is obtained as $(\hat{\theta}^*_{\alpha/2}, \hat{\theta}^*_{1 - \alpha/2} )$, where $\hat{\theta}^*_{\alpha/2}$ and  $\hat{\theta}^*_{1 - \alpha/2} $ denote the $\alpha/2$ and $1-\alpha/2$ quantiles of the bootstrap distribution, respectively. The PREB bootstrap is a semiparametric bootstrap method, as the marginal residuals $\bm{r}_t$ are computed using the fixed effect structure of model \eqref{eq:lograin_lmm}, while the day-level random effect predictors $\hat{u}_t^{sc}$ and residuals $\hat{\bm{e}}_{t}^s$ are generated nonparametrically. Importantly, it is an extension of the random effect block (REB) bootstrap originally used in the analysis of  \cite{chambersETAL2022} to accommodate clustered data with highly unbalanced cluster sizes. In the Oman trial dataset, the cluster (day) sizes $n_t$ range from $n_t = 1$ to $n_t = 58$ (see Figure \ref{fig:hist_nt} in Section \ref{sec:trial}), motivating the use of the PREB bootstrap. The unbalanced cluster sizes are handled through a combination of reflation in Step 2 and SRS and PPS sampling schemes in Steps 4 and 5 of Algorithm \ref{al:PREB}, respectively; see \cite{thoETAL2025} for more details.

\section{Application to Oman 2013 - 2018 Trial} \label{sec:realdata}
In this section, we apply our proposed attribution estimator to the Oman rainfall enhancement trial dataset from 2013 to 2018, introduced in Section \ref{sec:trial}, to assess the effectiveness of the ground-based ionization technology, and compare the results with those from the previous analysis of the same dataset using the estimator of \cite{chambersETAL2022}. 

Accordingly, we fit the linear mixed model \eqref{eq:lograin_lmm} to the log-transformed rainfall for the gauge-day observations located in the downwind corridors defined earlier. As the focus of this article is on comparing the newly proposed attribution estimator to that of \cite{chambersETAL2022}, rather than on selecting the optimal covariate set for modeling the log-rainfall, we follow \cite{chambersETAL2022} in determining the covariates to be used. Specifically, $\bm{x}_{it}$ contains the intercept term, the elevation of the rain gauge $i$, and the expected log-rainfall at gauge $i$ on day $t$. The expected log-rainfall values are obtained as fitted values from a preliminary linear mixed model fitted to the log-rainfall of upwind observations, using the meteorological variables from the AWS and Muscat International Airport described in Section \ref{sec:trial} as covariates; see Table 3 of \cite{chambersETAL2022} for details. The use of such a preliminary upwind model to capture the effect of meteorological processes on rainfall amounts that are independent of ionizer operation was developed in \cite{chambersETAL2012,chambersETAL2016}. While $\bm{x}_{it}$ is unrelated to the enhancement intervention, $\bm{z}_{it}$ is constructed to measure the exposure of each observation to ionization. It contains ten binary target indicators corresponding to the ten ionizers, as well as two interactions terms between elevation and the binary target indicators for the H1 and H2 ionizers. For example, the H1 target indicator equals one only if gauge $i$ is downwind of the H1 ionizer on day $i$ and the H1 ionizer was switched on on that day, and equals zero otherwise. 

\begin{table}[!tb]
\caption{REML estimates of fixed-effect coefficients and variance components from the linear mixed model \eqref{eq:lograin_lmm} fitted to the Oman trial data (2013--2018).}
\label{table:oman_reml}
\centering
\begin{tabular}{llr @{\hspace{1cm}} llr}
\toprule
\multicolumn{3}{c}{Fixed effects} & \multicolumn{2}{c}{Variance components} \\
\cmidrule(r){1-3} \cmidrule(l){4-6}
& Covariate & Estimate & & Component & Estimate \\
\midrule
\multirow{3}{*}{$\hat{\bmalpha}$} 
 & Intercept             & 0.280  & $\hat{\sigma}^2_u$ & Day random effect & 0.274 \\
 & Elevation             & -0.125 & $\hat{\sigma}^2_e$ & Error term & 1.856 \\
 & Expected log-rainfall & 0.856  &  &  \\
\addlinespace
\multirow{12}{*}{$\hat{\bmbeta}$} 
 & Target H1             & 0.289  &  &  \\
 & Target H2             & 0.258  &  &  \\
 & Target H3             & 0.238  &  &  \\
 & Target H4             & -0.153 &  &  \\
 & Target H5             & 0.433  &  &  \\
 & Target H6             & -0.203 &  &  \\
 & Target H7             & 0.222  &  &  \\
 & Target H8             & 0.059  &  &  \\
 & Target H9             & 0.481  &  &  \\
 & Target H10            & 0.034  &  &  \\
 & Elevation $\times$ Target H1 & -0.087 & & \\
 & Elevation $\times$ Target H2 & -0.217 & & \\
\bottomrule
\end{tabular}
\end{table}

Table \ref{table:oman_reml} presents the REML estimates for the fixed effect coefficients $\hat{\bmalpha}$ and $\hat{\bmbeta}$, and the variance components $\hat{\sigma}^2_u$ and $\hat{\sigma}^2_e$, which replicates the estimates reported in Table 4 of \cite{chambersETAL2022}. Unsurprisingly, the expected log-rainfall covariate has the largest effect. Most of the target indicators for the ionizers also exhibit positive effects on downwind log-rainfall,  with the exceptions of the H4 and H6 ionizers. The ratio $\hat{\sigma}^2_u / (\hat{\sigma}^2_e + \hat{\sigma}^2_u) = 0.129$ reflects the magnitude of day-to-day rainfall variability, indicating that the day random effect accounts for 12.9\% of the total variation in log-rainfall.  

\begin{figure}[!tb]
    \centering
    \includegraphics[width=1\linewidth]{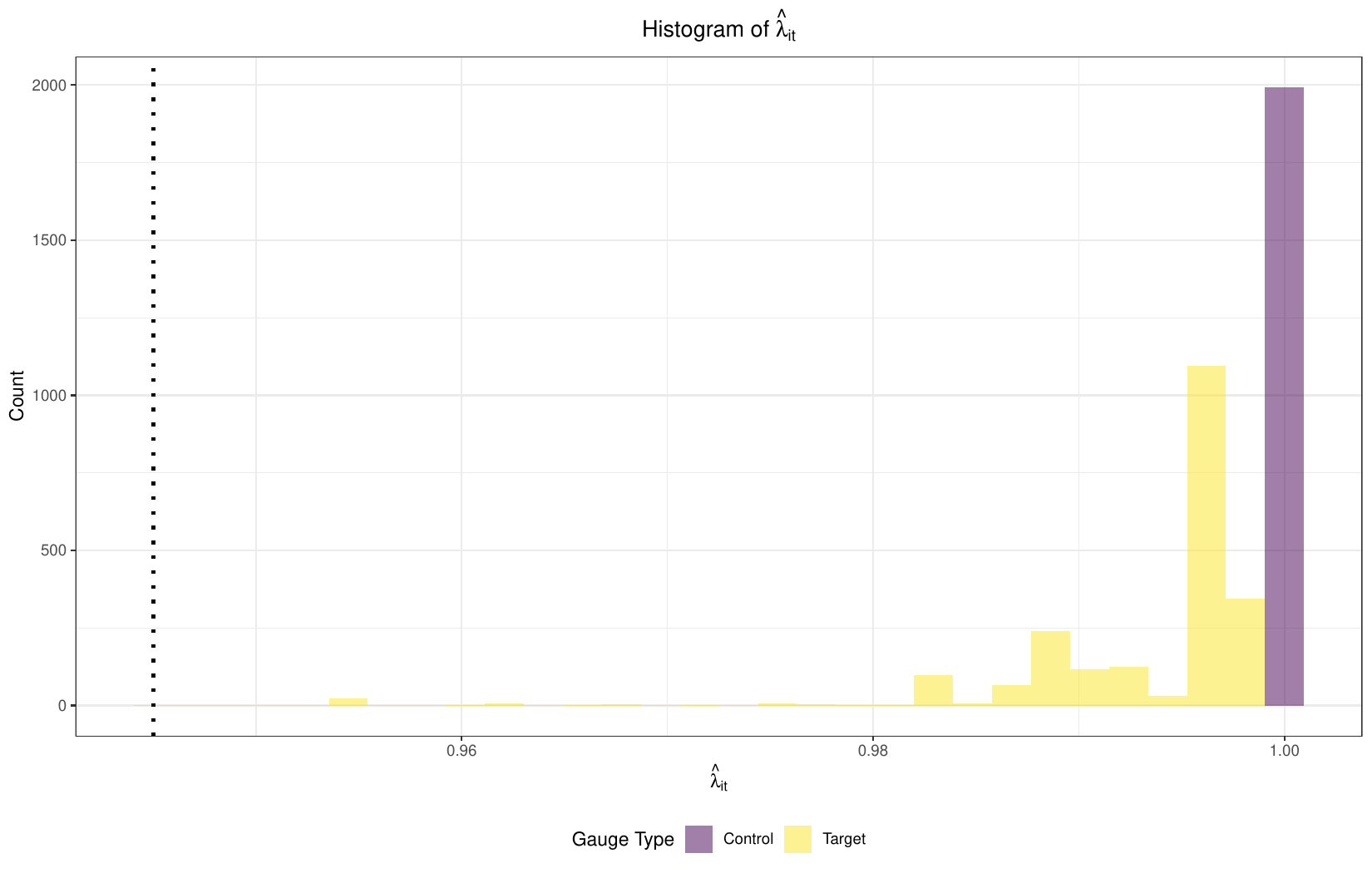}
    \caption{Histogram of bias-adjustment terms $\hat{\lambda}_{it}$ for all downwind observations, with yellow color representing the `target' observations and purple color representing the 'control' observations. The dotted line represents the alternative bias-adjustment term $\tilde{\lambda}$ from \cite{chambersETAL2022}. }
    \label{fig:hist_lambda_it}
\end{figure}

\begin{figure}[!tb]
    \centering
    \includegraphics[width=1\linewidth]{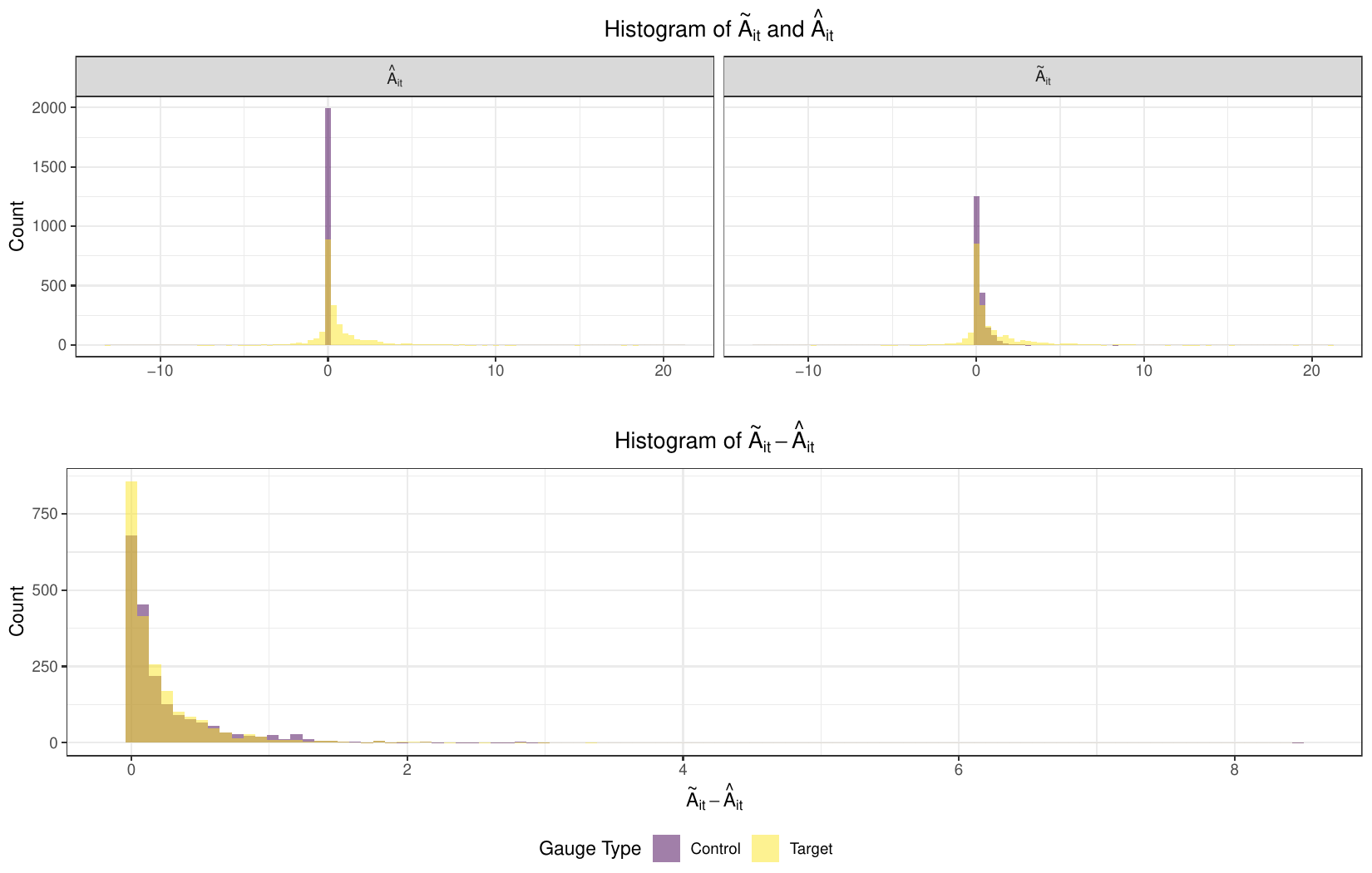}
    \caption{Histogram of $\hat{A}_{it}$ (top left), $\tilde{A}_{it}$ (top right) and $\tilde{A}_{it} - \hat{A}_{it}$ (bottom) for all downwind observations, with yellow color representing the `target' observations and purple color representing the 'control' observations. }
    \label{fig:hist_A_all}
\end{figure}

The parameter estimates in Table \ref{table:oman_reml} are used to compute the proposed bias-adjustment terms $\hat{\lambda}_{it}$ and the resulting attribution estimates $\hat{A}_{it}$ in Section \ref{sec:bias_adjustment_attribution} for all downwind observations. For comparison, we also compute the bias-adjustment term $\tilde{\lambda}$ proposed by \cite{chambersETAL2022} in equation \eqref{eq:tilde_lambda} and the corresponding attribution estimates $\tilde{A}_{it}$. Figure \ref{fig:hist_lambda_it} shows the distribution of $\hat{\lambda}_{it}$ together with the value of $\tilde{\lambda}$, while Figure \ref{fig:hist_A_all} presents the distribution of the attribution estimates $\hat{A}_{it}$ and $\tilde{A}_{it}$ as well as their differences. It can be seen that the adjustment term $\tilde{\lambda} = 0.945$ proposed by \cite{chambersETAL2022} is smaller than all values of $\hat{\lambda}_{it}$ for the downwind observations, which in turn leads to $\tilde{A}_{it} \geq \hat{A}_{it}$ for all observations as illustrated through the distribution of $\tilde{A}_{it} - \hat{A}_{it}$ that is entirely positive. Notably, Figures \ref{fig:hist_lambda_it} -- \ref{fig:hist_A_all} illustrate the coherence property of our proposed estimator, since $\hat{\lambda}_{it} = 1$ and $\hat{A}_{it} = 0$ for all `control' observations. In contrast, this property does not hold for the estimator of \cite{chambersETAL2022}, as $\tilde{A}_{it}$ are non-zero for the `control' observations. 

The overall attribution estimate obtained using our proposed method shows a $\hat{\theta} = 6.68\%$ increase in downwind rainfall attributable to the ionizer operation as a percentage of the total natural rainfall. In comparison, the overall attribution estimate based on the method of \cite{chambersETAL2022} is $\tilde{\theta} = 12.52\%$, which is substantially larger than our proposed estimate. This discrepancy arises directly from the different constructions of the bias-adjustment terms $\hat{\lambda}_{it}$ and $\tilde{\lambda}$, highlighting the importance of using theoretically grounded adjustment $\hat{\lambda}_{it}$ as discussed in Section \ref{sec:bias_adjustment_attribution}. We further investigate these differences empirically in Section \ref{sec:simulation} through a simulation study, where the results demonstrate that the proposed estimator $\hat{\theta}$ based on $\hat{\lambda}_{it}$ provides a more accurate estimate of the true attribution parameter $\theta$ than the alternative estimator $\tilde{\theta}$ that relies on $\tilde{\lambda}$.

\begin{figure}[!tb]
    \centering
    \includegraphics[width=1\linewidth]{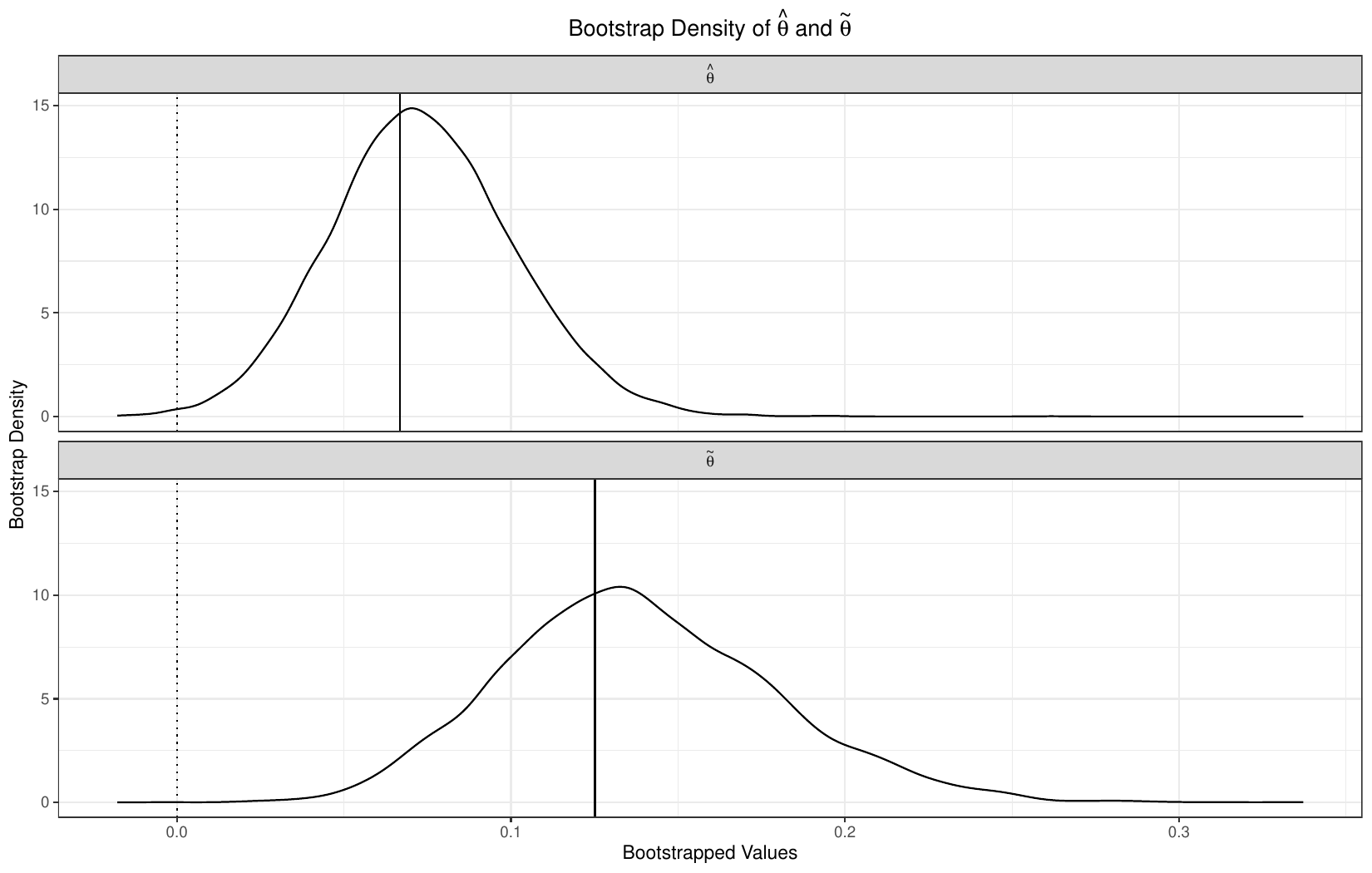}
    \caption{Bootstrap distributions of $\hat{\theta}$ (top) and $\tilde{\theta}$ (bottom) using the PREB bootstrap, with solid lines representing the corresponding point estimates from the original Oman data and dotted lines representing zero. }
    \label{fig:bootstrap_plot}
\end{figure}

As the number of downwind gauges with positive rainfall $n_t$ is highly unbalanced across trial days (see Figure \ref{fig:hist_nt}), we employ the PREB bootstrap in Section \ref{sec:PREB} with $B = 10000$ bootstrap replicates to obtain the bootstrap distribution of the proposed estimator $\{\hat{\theta}^{*(b)}: b=1,\cdots,B \}$, as well as that of \cite{chambersETAL2022}'s estimator $\{\tilde{\theta}^{*(b)}: b=1,\cdots,B \}$. The latter distribution is achieved by modifying Step 8 of Algorithm \ref{al:PREB} to compute the bias-adjustment term $\tilde{\lambda}^{*}$ based on the bootstrap sample data, along with the corresponding attributions $\tilde{A}_{it}^{*}$ and $\tilde{\theta}^{*}$. Figure \ref{fig:bootstrap_plot} presents these bootstrap distributions and Table \ref{table:bootstrap_summary_stat} provides summary statistics from these distributions, showing that the bootstrap distribution of $\tilde{\theta}$ is shifted to the right of that of $\hat{\theta}$. Consequently, the bootstrap 95\% confidence interval for the true overall attribution $\theta$ based on $\tilde{\theta}$ (6.64\% -- 22.39\%) is also shifted to the right relative to that based on $\hat{\theta}$ (2.21\% -- 12.88\%). Again, these shifts arise from the use of different bias-adjustment terms, i.e., $\tilde{\lambda}$ and $\hat{\lambda}_{it}$. The bootstrap $p$-values for testing a positive overall attribution, defined as the proportion of bootstrap estimates $\hat{\theta}$ or $\tilde{\theta}$ that are less than or equal to zero, are 0.0026 and 0.0001 based on $\hat{\theta}$ and $\tilde{\theta}$, respectively. Overall, bootstrap inference obtained under both the proposed estimator $\hat{\theta}$ and the alternative estimator $\tilde{\theta}$ indicates that the ionizer operation has a statistically significant positive effect on downwind rainfall at the 5\% significance level. However, the magnitude of this effect is consistently estimated to be smaller using the proposed estimator. A simulation study is conducted in the next section to further facilitate understanding of these differences.

\begin{table}[!tb]
\caption{Summary statistics for the bootstrap distributions of $\hat{\theta}$ and $\tilde{\theta}$ using the PREB bootstrap, where SD denotes standard deviation. }
\label{table:bootstrap_summary_stat}
\centering
\begin{tabular}{l rrrrrrrrr}
\toprule
& & \multicolumn{5}{c}{Percentile} & & &  \\
\cmidrule(r){3-7} 
& Min & 2.5\% & 25\% & 50\% & 75\% & 97.5\% &  Max & Mean & SD  \\
\midrule
$\hat{\theta}$ & -1.78\% & 2.21\% &  5.50\% &  7.26\% &  9.12\% & 12.88\% & 26.18\% &  7.36\% & 2.74\%  \\
$\tilde{\theta}$ & -0.31\% & 6.64\% & 11.02\% & 13.56\% & 16.44\% & 22.39\% & 33.72\% & 13.82\% & 4.02\% \\
\bottomrule
\end{tabular}
\end{table}

\section{Simulation Study}\label{sec:simulation}
We conduct a numerical study to better understand the differences in our proposed estimate $\hat{\theta}$ and \cite{chambersETAL2022}'s estimate $\tilde{\theta}$ observed in the previous section. Specifically, we simulate log-rainfall from model \eqref{eq:lograin_lmm} for the same $N = 4168$ downwind gauge-day observations with positive rainfall amounts. The true values of the parameters $\bmalpha, \bmbeta, \sigma^2_u$, and $\sigma^2_e$ are set to their estimated values from Table \ref{table:oman_reml} of Section \ref{sec:realdata}, while the covariate vectors $\bm{x}_{it}$ and $\bm{z}_{it}$ are fixed throughout the simulation. The random effects $u_i$ and error terms $e_{it}$ are regenerated for each of the $M = 500$ datasets. Based on the simulated $u_t$ and $e_{it}$, the true values of $A_{it}$ are computed using equations \eqref{eq:Ait} and \eqref{eq:Rit}, which then determine the true overall attribution $\theta$. For each simulated dataset, we apply our proposed method to obtain  $\hat{A}_{it}$ and $\hat{\theta}$, and we apply \cite{chambersETAL2022}'s method to obtain the corresponding $\tilde{A}_{it}$ and $\tilde{\theta}$. The PREB bootstrap is also applied with 1000 replicates to construct bootstrap 95\% confidence interval for $\theta$ using both $\hat{\theta}$ and $\tilde{\theta}$.

To assess point estimation performance, we compute the bias and mean squared errors (MSE) averaged over 500 simulations as $\textrm{Bias}(A_{it}) = 500^{-1} \sum_{m=1}^{500}N^{-1} \sum_{t=1}^{T} \sum_{i=1}^{n_t} ( \dot{A}_{it}^{[m]} - A_{it}^{[m]} )$ and $\textrm{MSE}(A_{it}) = 500^{-1} \sum_{m=1}^{500} N^{-1} \sum_{t=1}^{T} \sum_{i=1}^{n_t} ( \dot{A}_{it}^{[m]} - A_{it}^{[m]} )^2$ for individual attribution $A_{it}$, $\textrm{MSE}(\textrm{Average } A_{it}) =  500^{-1} \sum_{m=1}^{500}( N^{-1} \sum_{t=1}^{T} \sum_{i=1}^{n_t}\dot{A}_{it}^{[m]} -  N^{-1} \sum_{t=1}^{T} \sum_{i=1}^{n_t}A_{it}^{[m]} )^2$ for mean $A_{it}$, $\textrm{Bias}(\theta) = 500^{-1} \sum_{m=1}^{500} (\dot{\theta}^{[m]} - \theta^{[m]})$  and $\textrm{MSE}(\theta) = 500^{-1} \sum_{m=1}^{500} (\dot{\theta}^{[m]} - \theta^{[m]})^2$ for the overall attribution, where $\dot{A}_{it}$ and $\dot{\theta}$ generically denote estimates from either our proposed method or \cite{chambersETAL2022}'s approach, and the superscript $[m]$ denotes the $m$-th simulated dataset. For inferential performance, we compute the empirical coverage of the bootstrap 95\% confidence intervals for $\theta$ under both estimators.

\begin{figure}[!tb]
    \centering
    \includegraphics[width=1\linewidth]{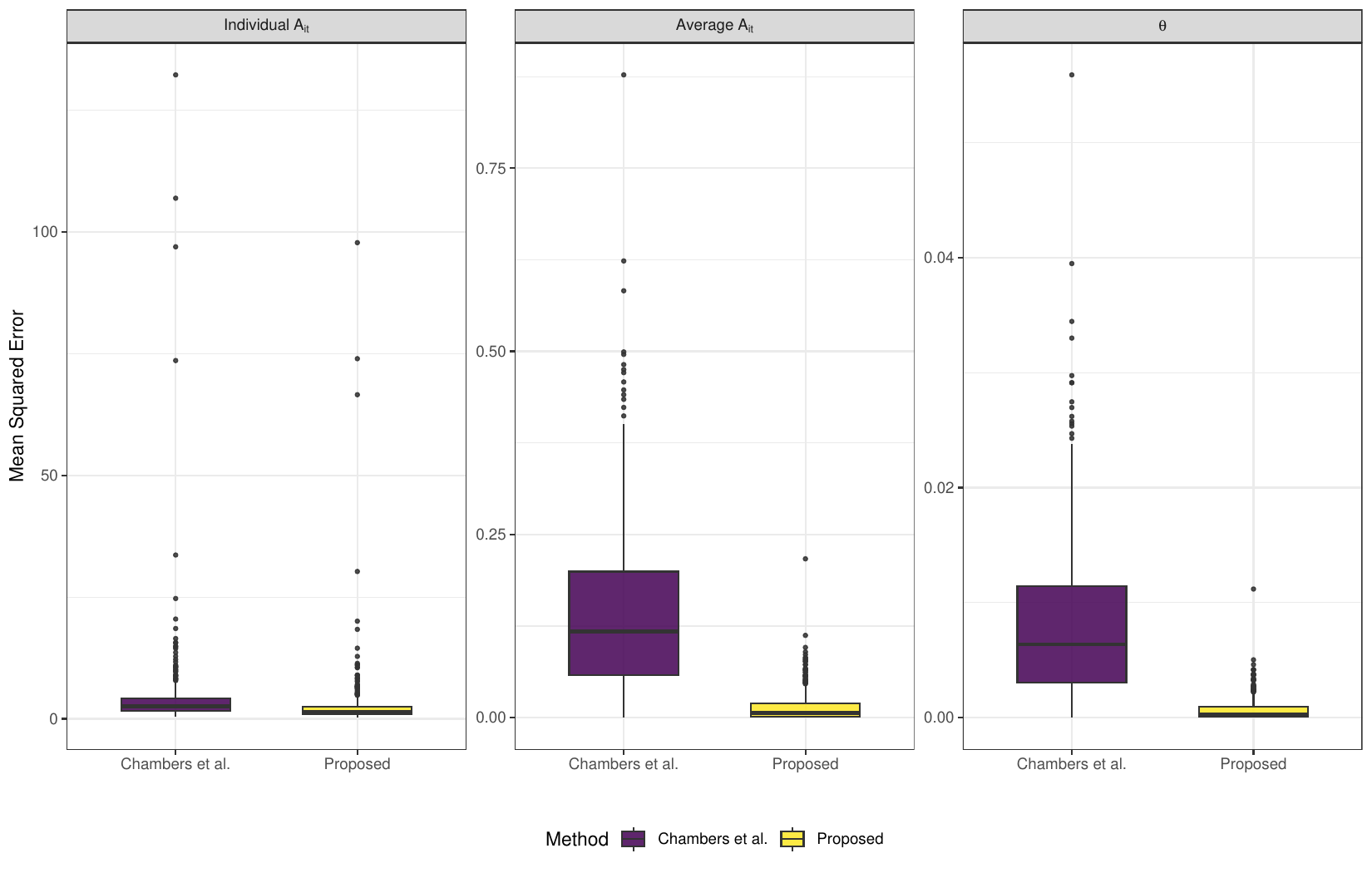}
    \caption{Boxplots summarizing MSEs for the proposed estimator and \cite{chambersETAL2022}'s estimator of individual $A_{it}$, mean $A_{it}$, and $\theta$ across 500 simulations.}
    \label{fig:mse_plot}
\end{figure}

\begin{figure}[!tb]
    \centering
    \includegraphics[width=1\linewidth]{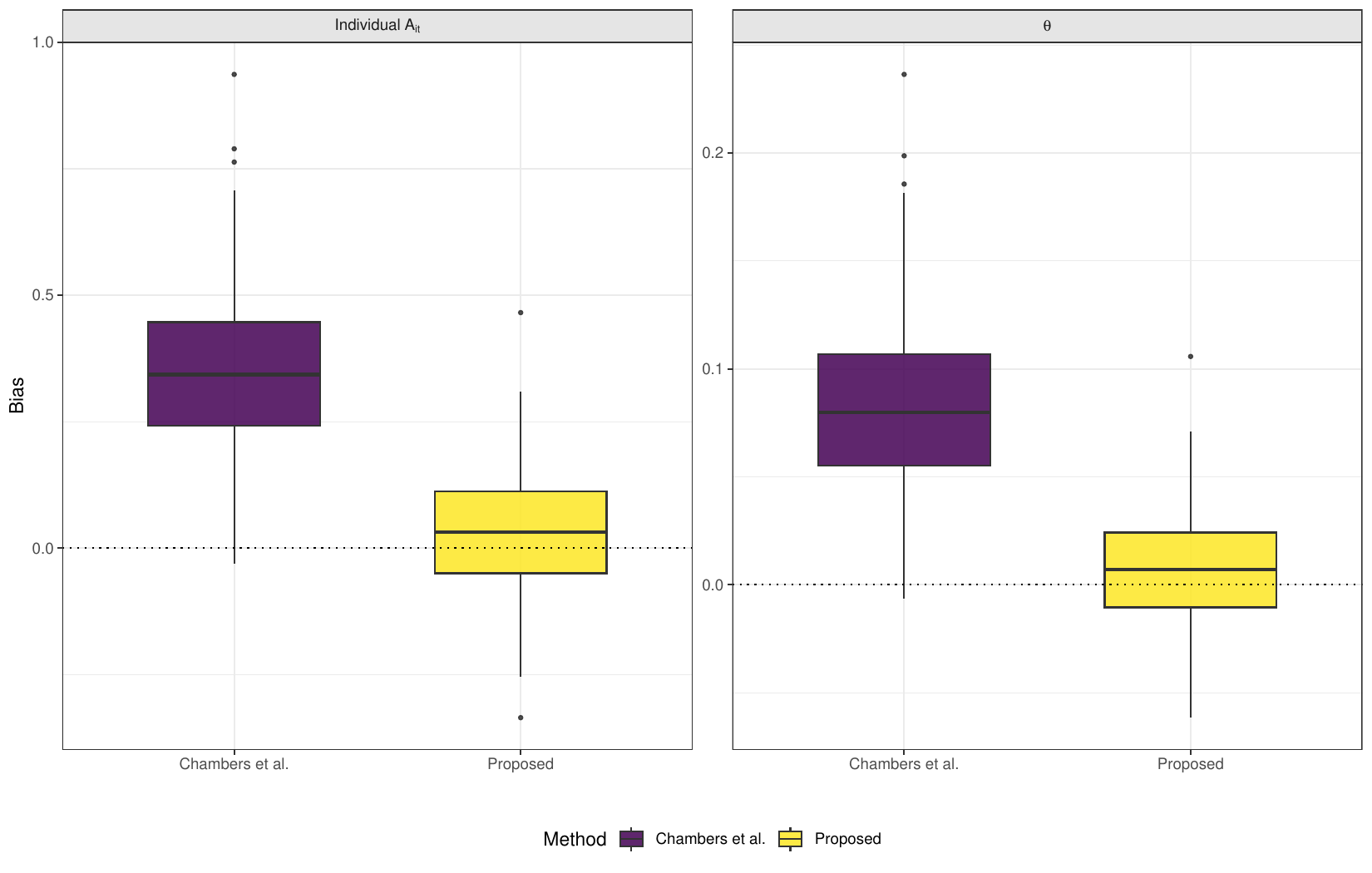}
    \caption{Boxplots summarizing biases for the proposed estimator and \cite{chambersETAL2022}'s estimator of individual $A_{it}$ and $\theta$ across 500 simulations.}
    \label{fig:bias_plot}
\end{figure}

Table \ref{table:mse_bias_simulation} demonstrates that our proposed method substantially outperforms the approach of \cite{chambersETAL2022} in estimating individual ${A}_{it}$, mean $A_{it}$, and the overall attribution $\theta$, as indicated by consistently smaller averaged MSEs. This is further supported by the boxplots in Figure \ref{fig:mse_plot}, which display the MSEs across 500 simulated datasets: the MSEs from our proposed method are consistently lower, although for individual $A_{it}$ the difference is less visible due to outliers stretching the y-axis scale. Moreover, the bias of our proposed $\hat{A}_{it}$ ($0.032$) is substantially smaller than that of $\tilde{A}_{it}$ ($0.348$), which is consistent with the finding from the Oman trial analysis that $\tilde{A}_{it} \geq \hat{A}_{it}$ for all observations. Table \ref{table:mse_bias_simulation} also shows that the bias of $\hat{\theta}$ is negligible ($0.76\%$), while $\tilde{\theta}$ exhibits a large positive bias ($8.22\%$). This aligns with the observation in Section \ref{sec:realdata} that $\tilde{\theta}$ was considerably larger than $\hat{\theta}$. Figure \ref{fig:bias_plot} further illustrates that $\hat{A}_{it}$ and $\hat{\theta}$ are unbiased, while $\tilde{A}_{it}$ and $\tilde{\theta}$ exhibit clear positive biases.

\begin{table}[!tb]
\caption{Bias, Mean square error (MSE), and coverage for the proposed estimator and \cite{chambersETAL2022}'s estimator of individual $A_{it}$, mean $A_{it}$, and $\theta$ based on the simulation study.}
\label{table:mse_bias_simulation}
\centering
\begin{tabular*}{\linewidth}{l rrrrrrrr}
\toprule
& $\textrm{Bias}(A_{it})$ &$\textrm{MSE}(A_{it})$ & $\textrm{MSE}(\textrm{Average } A_{it})$ & $\textrm{Bias}(\theta)$ & $\textrm{MSE}(\theta)$ & Coverage \\
\midrule
Proposed & $0.032$ & $2.653$ & $0.014$  & $0.76\%$ & $0.001$ & $96.00\%$ \\ 
Chambers et al. & $0.348$ & $4.256$ & $0.144$  & $8.22\%$ & $0.008$ & $22.60\%$ \\ 
\bottomrule
\end{tabular*}
\end{table}

\begin{figure}[!tb]
    \centering
    \includegraphics[width=1\linewidth]{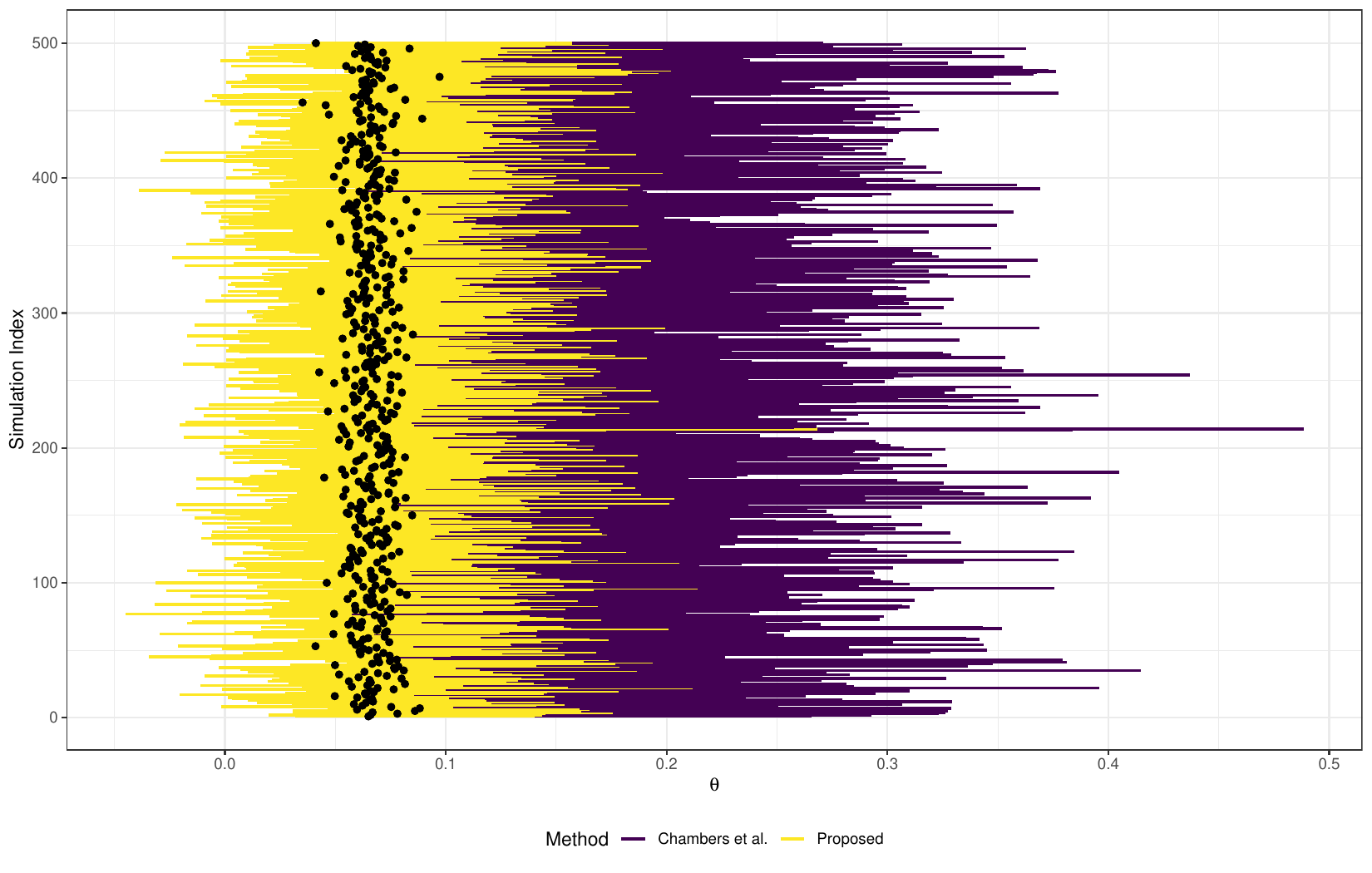}
    \caption{Bootstrap 95\% confidence intervals for $\theta$ across simulated datasets (y-axis) based on the proposed estimator (yellow) and \cite{chambersETAL2022}'s estimator (purple). Black points denote the true $\theta$.}
    \label{fig:coverage_plot}
\end{figure}

Regarding inferential performance, the bootstrap 95\% confidence intervals based on $\hat{\theta}$ achieve 96\% empirical coverage, close to the nominal level, whereas intervals based on $\tilde{\theta}$ suffer from severe undercoverage. Figure \ref{fig:coverage_plot} provides a detailed view across all 500 simulations, showing that the intervals using our proposed estimator reliably cover the true $\theta$, while the intervals using \cite{chambersETAL2022}'s estimator frequently fail due to the positive bias of $\tilde{\theta}$ shifting intervals to the right. This pattern aligns with the rightward shift of the bootstrap distribution of $\tilde{\theta}$ observed in Section \ref{sec:realdata}.

To summarize, the proposed estimators $\hat{A}_{it}$ and $\hat{\theta}$ consistently outperform the corresponding estimates from \cite{chambersETAL2022} in terms of point estimation. Moreover, the bootstrap intervals based on $\hat{\theta}$ exhibit better coverage. Combined with the consistency of these findings with the real data analysis in Section \ref{sec:realdata}, the results suggest that the proposed method provides a more reliable and theoretically grounded approach for the Oman rainfall enhancement trial.

\section{Conclusion}\label{sec:conclusion}
We propose a new estimator for attribution, a measure of the additional raw-scale rainfall attributable to the operation of rainfall enhancement technology in model-based analyses of rainfall enhancement trial data. Rainfall measurements are often log-transformed prior to fitting linear mixed models to satisfy normality assumptions, which can induce bias when back-transforming log-rainfall to the original scale in estimating attribution quantities. We derive an approximation for this bias and incorporate the resulting adjustment term into the proposed estimator to correct for it. The proposed estimator also satisfies a coherence property: observations that do not receive any rainfall enhancement intervention are guaranteed to have zero estimated attribution. We further employ a proportional random effect block bootstrap to conduct statistical inference on the attribution quantities. Applying both the proposed estimator and the alternative estimator by \cite{chambersETAL2022} to the Oman rainfall enhancement trial data from 2013 to 2018, the proposed method estimates a 6.68\% increase in downwind rainfall attributable to the ground-based ionizer operation as a percentage of the total natural rainfall, compared with 12.52\% from the alternative estimator. Bootstrap inference based on both estimators indicates a statistically significant positive effect of the ionization technology on downwind rainfall at the 5\% significance level, although the bootstrap distribution associated with the alternative estimator is shifted to the right of that of the proposed estimator. The simulation study further shows that the proposed estimator greatly outperforms the alternative estimator in terms of point estimation accuracy for the attribution quantities, and that the bootstrap confidence interval based on the proposed estimator achieves coverage close to the nominal level, whereas the interval based on the alternative estimator suffers from undercoverage. Importantly, these simulation findings are consistent with the differences observed in the Oman trial analysis, providing empirical support for the conclusions drawn from the proposed method. 

A logical next step is to extend the proposed estimator to accommodate spatial linear mixed models that allow for spatially correlated random effects, where such spatial correlation could be modelled through spatial covariance functions e.g., the  Mat\'{e}rn covariance function \citep{Matern1960}. This would be useful as rainfall amounts tend to be spatially correlated due to the localized nature of precipitation. It would also be interesting to investigate the effect of model misspecification on the estimation performance of the proposed estimator, such as misspecified $\bm{x}_{it}$ covariate vector through omitted variables. 
Finally, while this article has focused on the use of the PREB bootstrap designed specifically for linear mixed models, more general bootstrap methods for clustered data such as the cluster bootstrap \citep{davisonANDhinkley1997} could be considered for inference of the attribution quantities. 

\backmatter



\bmhead{Acknowledgements}

Zhi Yang Tho was supported by the Handbury Foundation Project.


\section*{Declarations}
\bmhead{Conflict of interest} The authors declare no conflict of interest.










\bibliography{bibliography.bib}

\end{document}